\pdfoutput=1
\documentclass[twocolumn,twocolappendix,acknowledgments]{aastex631}

\usepackage{graphicx}
\usepackage{bm}	
\usepackage{booktabs}
\usepackage{threeparttable}
\usepackage{multirow}
\usepackage{soul}
\usepackage{amsmath}
\shorttitle{satellite-halo connection}
\shortauthors{Zhou \& Han}
\graphicspath{{./}{figure/}}
\begin{document}
\title{Mining the information content of member galaxies in the halo mass modelling}

\author{Yanrui Zhou}
\affiliation{Department of Astronomy, School of Physics and Astronomy, Shanghai Jiao Tong University, Shanghai 200240, China\\}
\affiliation{ Key Laboratory for Particle Astrophysics and Cosmology (MOE), Shanghai 200240, China\\}
\affiliation{Shanghai Key Laboratory for Particle Physics and Cosmology, Shanghai 200240, China\\}

\author[0000-0002-8010-6715]{Jiaxin Han}%\email{jiaxin.han@sjtu.edu.cn}
\affiliation{Department of Astronomy, School of Physics and Astronomy, Shanghai Jiao Tong University, Shanghai 200240, China\\}
\affiliation{ Key Laboratory for Particle Astrophysics and Cosmology (MOE), Shanghai 200240, China\\}
\affiliation{Shanghai Key Laboratory for Particle Physics and Cosmology, Shanghai 200240, China\\}

\correspondingauthor{Jiaxin Han}
\email{123zyr@sjtu.edu.cn (YZ), jiaxin.han@sjtu.edu.cn (JH)}

% \author[0000-0002-0786-7307]{Greg J. Schwarz}
% \affiliation{American Astronomical Society \\
% 1667 K Street NW, Suite 800 \\
% Washington, DC 20006, USA}

% \author{August Muench}
% \affiliation{American Astronomical Society \\
% 1667 K Street NW, Suite 800 \\
% Washington, DC 20006, USA}

% \collaboration{6}{(AAS Journals Data Editors)}

% \author{Butler Burton}
% \affiliation{Leiden University}
% \affiliation{AAS Journals Associate Editor-in-Chief}

% \author{Amy Hendrickson}
% \altaffiliation{AASTeX v6+ programmer}
% \affiliation{TeXnology Inc.}

% \author{Julie Steffen}
% \affiliation{AAS Director of Publishing}
% \affiliation{American Astronomical Society \\
% 1667 K Street NW, Suite 800 \\
% Washington, DC 20006, USA}

% \author{Magaret Donnelly}
% \affiliation{IOP Publishing, Washington, DC 20005}



\begin{abstract}
Motivated by previous findings that the magnitude gap between certain satellite galaxy and the central galaxy can be used to improve the estimation of halo mass, we carry out a systematic study of the information content of different member galaxies in the modelling of the host halo mass using a machine learning approach.
%It has been suggested that the inclusion of certain magnitude gaps between satellites and the central galaxy can substantially improve the halo mass estimation based on the central magnitude. In this work, we investigate the contributions of different member galaxies to the improvement with machine learning. 
We employ data from the hydrodynamical simulation IllustrisTNG and train a Random Forest (RF) algorithm to predict a halo mass from the stellar masses of its member galaxies. Exhaustive feature selection is adopted to disentangle the importances of different galaxy members. We confirm that an additional satellite does improve the halo mass estimation compared to that estimated by the central alone. However, the magnitude of this improvement does not differ significantly using different satellite galaxies. When three galaxies are used in the halo mass prediction, the best combination is always that of the central galaxy with the most massive satellite and the smallest satellite. Furthermore, among the top 7 galaxies, the combination of a central galaxy and two or three satellite galaxies gives a near-optimal estimation of halo mass, and further addition of galaxies does not raise the precision of the prediction. We demonstrate that these dependences can be understood from the shape variation of the conditional satellite distribution, with different member galaxies accounting for distinct halo-dependent features in different parts of the cumulative stellar mass function.
\end{abstract}


% \keywords{Classical Novae (251) --- Ultraviolet astronomy(1736) --- History of astronomy(1868) --- Interdisciplinary astronomy(804)}

\section{Introduction}
According to the standard cosmological paradigm, galaxies are believed to form and evolve in dark matter halos\citep[]{white&ree1978}. Hence the properties of galaxies are tightly linked to the properties of their dark haloes, giving rise to the so-called the galaxy-halo connection. Studying this connection is of great value in understanding the mechanisms of galaxy formation and evolution, while also providing a way to infer the properties of dark matter haloes through galaxy observations. %(see \citet{W&Tinker2018} for a review).

%As the halo mass is a fundamental property of a halo, its precise estimation in observations is essential for quantifying the galaxy-halo connection. % and for issues regarding halo assembly bias. In addition to direct measurements through gravitational lensing\citep[e.g.][]{Mandelbaum, Han14, Wang21} or dynamical modelling~\citep[e.g.][]{Han16,Li21}, various indirect approaches have also been developed to infer the halo mass statistically. These include the abundance matching(AM)\citep[e.g.][]{Kravtsov1999,Moustakas2002,Tasitsiomi2004,Vale2004,Yang2005,Yang2007,Conroy2006,Moster2010}, halo occupation distribution (HOD)\citep[e.g.][]{Jing1998,Berlind2002,Cooray2002,Zheng2005} and conditional luminosity function(CLF)\citep[e.g.][]{Yang2003,vandenBosch2002,Cooray2006}.

One of the main characterisations of the galaxy-halo connection is the stellar mass-halo mass relation \citep[SHMR; see e.g.,][for a review]{W&Tinker2018}.
For central galaxies, this SHMR has been much discussed and well constrained observationally. However, a non-negligible scatter exists in the relation, which may be attributed to the varying assembly histories of galaxies and their different large-scale environments\citep[e.g.][]{Behroozi2010,Guo2010,Leauthaud2012,Reddick2013,GoldenMarx2018,GoldenMarx2019,Feldmann2019,Bradshaw2020}.

Some recent studies have shown that the magnitude or mass difference (a.k.a. gap) between the central galaxy and some satellite galaxies may contain information about the assembly history of their host halo~\citep[e.g.][]{Harrison2012,Deason2013,Solanes2016,Kang2016} and thus could be used to tighten the SHMR. 
The magnitude gap between the brightest central galaxy (BCG) and the second brightest galaxy (M12) was originally studied as a diagnosis for selecting fossil groups~\citep[]{Ponman1994,Jones2003,Sales2007,vonBenda-Beckmann2008}.
Later, \citet{Dariush2010} and \citet{Tavasoli2011} proposed that the magnitude gap between the BCG and the fourth brightest galaxy (M14) is a better indicator for selecting fossil groups. %, and M14 has become popular in studies since then when describing the magnitude gap\citep[e.g.][]{GoldenMarx2018,GoldenMarx2019,Golden-Marx2021,Harrison2012}. 
Subsequent studies have shown that using these gaps, in addition to the central luminosity, can indeed substantially reduce the scatter in the halo mass estimation~\citep[e.g.][]{More2012,Hearin2013,Shen2014,Lu2015,GoldenMarx2018,GoldenMarx2019,Golden-Marx2021,Wang21a}.
However, it is still not known whether the improvement in the halo mass estimation is equally effective for gaps between BCG and different ranked satellite galaxies, and which gap can optimally constrain the dark halo mass, or equivalently, which satellite galaxy provides the most information in constraining the halo mass.

A related question is how many satellites are needed to optimally constrain the halo mass.
%In general, satellite galaxies can be used as tracers dark matter in a halo, and the total stellar mass can be used as an indicator of halo mass\citep[e.g.]{Zaritsky1997,Prada2003,Conroy2007}. 
Taking information from all member galaxies would certainly improve the halo mass constraint. For example, as satellite galaxies are expected to trace the dark matter distribution in a halo~\citep[e.g.][]{Han16}, the total stellar mass or total luminosity can be used as a good proxy for halo mass~\citep[e.g.][]{Zaritsky1997,Prada2003,Yang2005,Conroy2007,Han15,Wang21}. However, the complete population of all member galaxies are usually not available observationally. 
\citet{Bradshaw2020} demonstrated that employing the sum of the stellar mass of the central and the $N$ most massive satellites (${\rm cen}+N$) as a new halo mass estimator can effectively reduce the scatter compared to using only the stellar mass of the central galaxy ($M_{*,\mathrm{cen}}$).
They also showed that the scatter adopting this estimator already approaches that using the total stellar mass of all galaxy members ($M_{*,\mathrm{tot}}$). However, it should be noted that the total stellar mass alone could miss information from the relative mass distribution of different satellites, and thus may not be optimal itself in estimating the halo mass.
Thus it remains to be seen which of the satellite galaxies play a major role and how to best combine the galaxy information to maximize the accuracy of halo mass prediction when only a few satellite galaxies are available.

In this work, we seek to clarify the roles played by different satellites in the estimation of the halo mass. The data we employed is from the hydrodynamic simulation IllustrisTNG. We make use of a machine learning technique called Random Forest (RF) regression to model the nonlinear joint connections between halo mass and the first few satellites, while sorting out the relative importances of different satellite combinations using the exhaustive feature selection method. 
We find that there is indeed an optimal combination of satellites that can lead to a nearly saturated improvement in the halo mass constraint. We further examine the results in the context of the conditional satellite distribution, showing that the optimal satellite combination can be clearly mapped to distinct features in the conditional mass function.

The paper is organized as follows:
In section \ref{sec:data} we introduce the IllustrisTNG simulation on which our analysis is based, and the process of data processing and filtering.
In section \ref{sec:method} we describe the machine learning method and the training process. The main results from the machine learning analysis are presented in section \ref{sec:result}. In section \ref{sec:CMF} we examine the results in the context of the conditional satellite distribution. Summary and conclusions are presented in section \ref{sec:summary}.

\section{Data}
\label{sec:data}
\subsection{IllustrisTNG}
Our analysis is based on data from the IllustrisTNG,\footnote{\url{https://www.tng-project.org/}} a suite of state-of-the-art magnetohydrodynamical cosmological simulations~\citep[]{Naiman2018_TNG,Springel2018_TNG,Pillepich2018_TNG,Nelson2018_TNG,Marinacci2018_TNG} run with the moving-mesh code \textsc{Arepo}~\citep[]{Springel_2010}. %, and self-gravity is computed with the Tree-PM approach. 
TNG is the successor of the original Illustris simulation, while improving in many aspects of its galaxy formation recipes. %including AGN feedback, cooling star-formation and supernovae feedback.
TNG follows the $\Lambda$ Cold Dark Matter cosmology adopting parameters from the Planck observations~\citep[]{Planck2016}, with $\Omega_\Lambda=0.6911$, $\Omega_m=0.3089$, $\Omega_b=0.0486$, $\sigma_8=0.8159$ and $h=0.6774$. %The specific cosmological parameter values are shown in the table\ref{Tab:cosmological parameters}.
% \begin{table}
% 	\centering
% 	\caption{Cosmological parameters of IllustrisTNG. From left to right: cosmological constant, matter density, baryonic density, normalisation, spectral index and Hubble constant.}
% 	\label{Tab:cosmological parameters}
% 	\begin{tabular}{cccccc} % four columns, alignment for each
% 		\hline
%         $\Omega_{\Lambda,0}$ &$\Omega_{m,0}$ & $\Omega_{b,0}$& $\sigma_8$ &$n_s$& h\\
%         \hline
%         0.6911 &0.3089& 0.0486&0.8159 & 0.9667 &0.6774\\
% 		\hline
% 	\end{tabular}
% \end{table}
The full TNG suit consist of simulations run in three different boxsizes of roughly 50, 100 and 300 $\mathrm{Mpc}$, referred to as TNG50, TNG100 and TNG300 respectively, each of which are also run at three or four levels of resolutions. In this work, we choose data from the highest resolution runs of TNG100 and TNG300 (named TNG100-1 and TNG300-1 in the data release), as TNG300 has the largest volume and therefore provides a large sample, while TNG100 has a higher resolution compared to TNG300. Given that the cosmologies are the same for both simulations, the data from the two are joined together in order to cover a larger halo mass range. More details on TNG100 and TNG300 are provided in Table \ref{tab:TNG300&TNG100 parameters}.
\begin{table}
	\centering
	\caption{Parameters of the TNG100-1 and TNG300-1. From left to right: side length of the simulation box, the number of dark matter particles, and the masses of dark matter and baryonic particles.}
	\label{tab:TNG300&TNG100 parameters}
	\begin{tabular}{cc*{5}c} % four columns, alignment for each
	    \hline
         Simulation & $L_{\rm box}[\mathrm{Mpc}]$ & $N_{\rm DM}$ &$m_{\rm DM}[\rm M_\odot]$ &$m_{\rm b}[\rm M_\odot]$\\
        \hline
         TNG100-1 & 110.7& $1820^3$&$7.5\times10^6$&$1.4\times10^6$\\
        \hline
         TNG300-1 & 302.6& $2500^3$&$5.9\times10^7$&$1.1\times10^7$\\
        \hline
	\end{tabular}
\end{table}
\subsection{The halo and galaxy sample}
Our study focuses on the sample at redshift $z = 0$. The halo mass is defined as $M_{\rm 200mean}$, the total mass in a sphere around the halo centre with an enclosed density of 200 times the mean density of the universe. We define satellite galaxies as those located within the virial radius, $R_{\rm 200mean}$, of the host halo except the central galaxy. %and drop the ones with only dark matter components without baryonic matter. 
As the stellar mass function of the simulations becomes incomplete at $\sim 10^6 \rm M_\odot$ ($\sim 10^7 \rm M_\odot$) for TNG100 (TNG300), we only consider galaxies above these stellar mass limits respectively. 

\begin{figure}[t]
  \centering
  \includegraphics[width=\columnwidth]{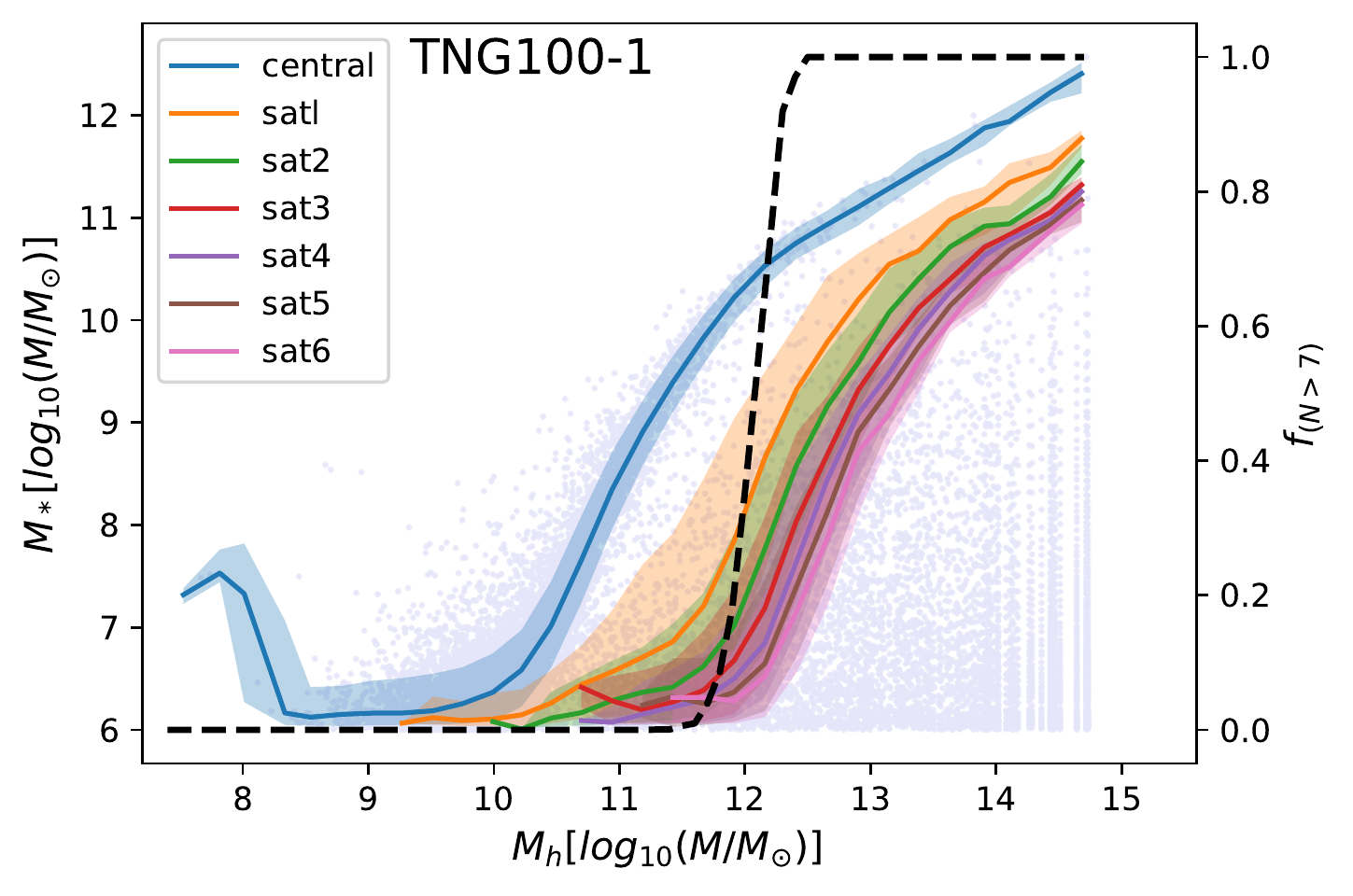}
  \hspace{1cm}
  \includegraphics[width=\columnwidth]{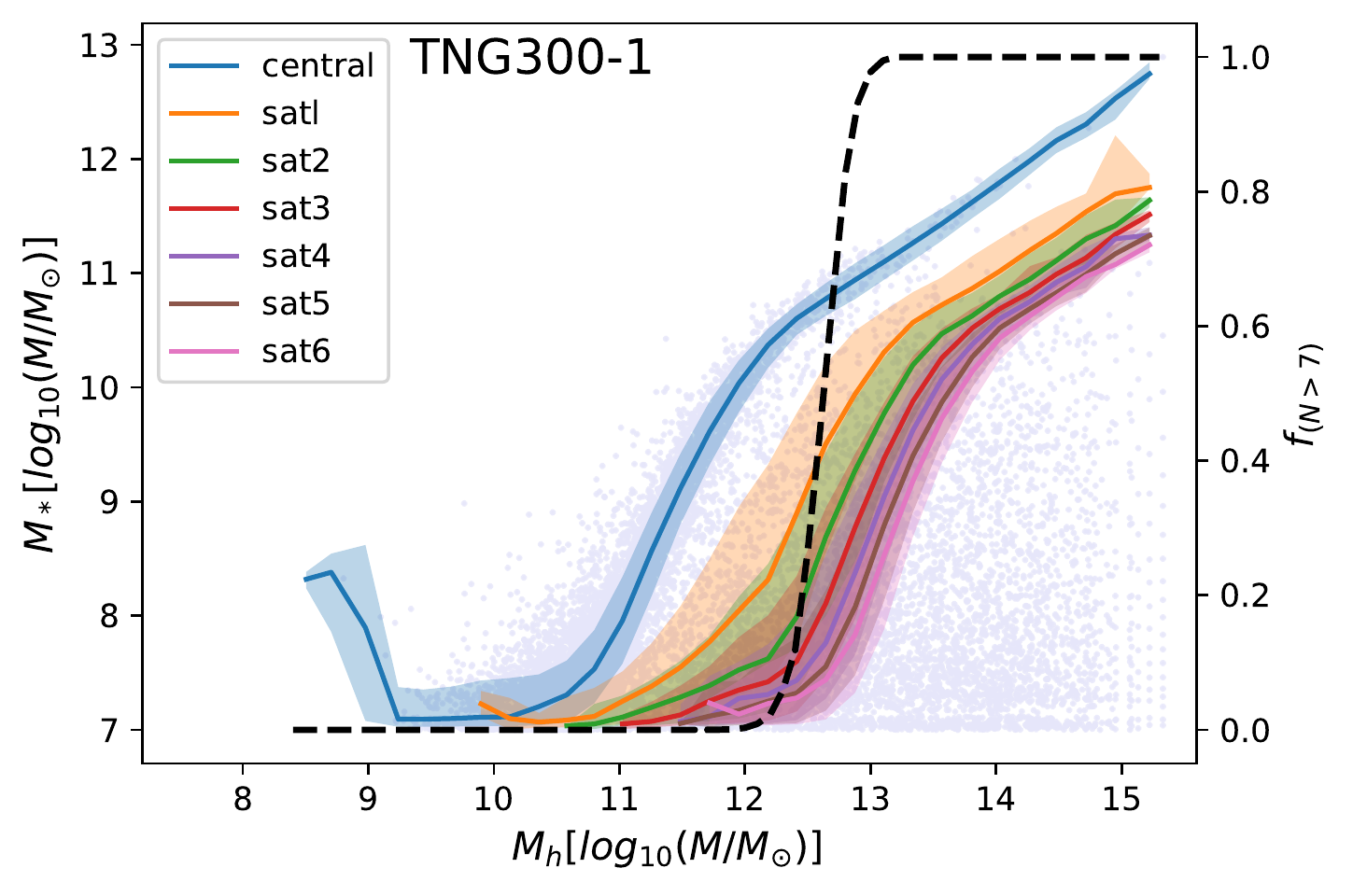}
  \caption{Stellar mass distributions of member galaxies in TNG100 (upper panel) and TNG300 (bottom panel). Each coloured solid curve shows the median stellar mass to halo mass relation for galaxies of a given rank as labelled, while the corresponding shaded region is bounded by the $16^{\rm th}$ and $84^{\rm th}$ percentiles in the stellar mass distribution. The light purple points in the background show the distribution of all the member galaxies. The black dashed line shows the fraction of halos with more than 7 members as a function of halo mass. }
  \label{fig:completeness analysis}
\end{figure}

We further demand that each halo contains at least 7 member galaxies in the halo mass range studied. As lower mass halos are typically resolved with fewer number of satellites, this richness cut translates to a cut in the halo mass for our sample. In Fig.~\ref{fig:completeness analysis} we plot the stellar mass-halo mass relation for the top 7 most massive galaxies in each halo, as well as the fraction of halos with more than 7 members as a function of halo mass.
As can be seen from Fig.~\ref{fig:completeness analysis}, limiting the halo mass to $M>10^{12.3}\rm M_{\odot}$ in TNG100 ensures that the top 7 galaxies are well resolved with stellar masses above $10^6 \rm M_\odot$.  For the TNG300 sample, a corresponding halo mass limit of $10^{12.8}\rm M_{\odot}$ can be found.
Under these selection criteria, we are left with 1235 valid halos in TNG100 and 8413 halos in TNG300, spanning a combined mass range of  $10^{12.3} < M_{\rm halo}/\rm M_{\odot} \lesssim 10^{15.3}$.

It is known that the galaxy properties do not fully converge between TNG100 and TNG300 due to the resolution dependence of the hydrodynamical solver used. To correct this, we simply multiply the stellar masses in TNG300 by a constant factor of 1.4 before combining it with TNG100, following \citet{Pillepich2018_TNG}.

% \begin{figure*}
%   \centering
%   \includegraphics[width=\columnwidth]{TNG100_300_centralhm_compare.pdf}
%   \includegraphics[width=\columnwidth]{/TNG100_300_centralhm_correct.pdf}
%   \caption[SHMR of TNG100 and TNG300]{SHMR of TNG100 and TNG300 for central galaxies, solid curves stand for the median stellar to halo mass relation, shaded regions correspond to $16^{th}$ and $84^{th}$ percentiles.Left panel: Original SHMR of TNG100 and TNG300, where it can be observed that the halo mass of TNG100 is slightly lower than that of TNG300 at fixed stellar mass. Right panel: Rescaled SHMR of TNG100 and TNG300, TNG100 has been corrected to TNG100*, in the range halo mass \textgreater$10^{12}\rm M_{\odot}$, comparing the two curves, it can be seen that they are now basically consistent.}
%   \label{fig:SHMR correct}
% \end{figure*}

\section{Method}
\label{sec:method}
%In recent decades, machine learning has been widely used for its powerful computing and predictive capabilities. It can be used to automate feature learning and modelling to help discover features that cannot be explicitly refined, as well as to explore the hidden structure and correlation of complex high-dimensional data, and help uncover unknown objects and physical properties. Based on these advantages, machine learning methods have been utilised increasingly in cosmology studies.
%We aim to explore the contribution of satellite galaxies to the prediction of halo mass and to find whether there exists a remarkable satellite galaxy that plays a significant role in the prediction process. To carry out a systematic study of this problem, Machine Learning is a suitable way of proceeding. 
We employ the Random Forest (RF) algorithm~\citep[]{Breiman2001_RF} as implemented in \textsc{scikit-learn}~\citep[]{scikit-learn} to analyze the relation between halo mass and the masses of member galaxies.
RF is a supervised machine learning algorithm that can be trained to map out the relation between the input and output data in a non-parametric way. It is an ensemble method that aggregates many base estimators called decision trees via the bagging approach. This enables the RF to overcome the common problem of overfitting faced by a single decision tree and improve the generalization ability of the model.
Due to its simplicity and efficiency, RF has been widely used in many recent studies in astrophysics~\citep[e.g.,][]{Hoyle2015,Man2019,Petulante2021,Shi22}. In the following we explain the algorithm in more detail. %Shi2021,Li2022}

\subsection{Decision tree}
As the basic unit of a RF, a decision tree is a tree-like decision model. % as the name implies. 
For a given input parameter space, a decision tree aims at partitioning the parameter space into multiple nodes such that each node is mapped to a single prediction. The partitioning is done by splitting the parameter space along one dimension at a time according to a certain criterion, forming a tree-like structure after multiple operations. The complete input parameter space forms the root node of the tree, while nodes that no longer split are called leaf nodes. The predictions in the leaf nodes can be either discrete classes or continuous values, corresponding to a classification or a regression tree. For this analysis we use regression trees. 

Consider a given data set consisting of $n$ observations,  $D=\{(\vec{x}_1,y_1),(\vec{x}_2,y_2),...,(\vec{x}_i,y_i),...(\vec{x}_n,y_n)\}$,
where $\vec{x}_i$ is an $m$-dimensional vector with $m$ input features, $y_i$ is the target feature we want to predict, and $i=1..n$ represents $n$ observations. Expressing the decision tree as a function $f(\vec{x})$, the goal of the regression is to find a $f(\vec{x})$ that minimizes the Mean Squared Error (MSE) of the data set (also referred to as impurity)
\begin{align}
  \mathrm{MSE}= \frac{1}{n}\sum_{i=1}^{n}(y_i-f(\vec{x}_i))^2.
  \label{eq:MSE}
\end{align}
This is achieved by choosing an appropriate division at each step to minimize the MSE in each node, with $f(\vec{x})$ replaced by the mean value of $y$ in the node.
%To minimize the overall MSE, the MSE of each leaf needs to be minimized. Hence, when choosing splitting variable and splitting point in each division, the decision tree will select the division that minimizes the MSE of the nodes.

More specifically, starting from the root node, we recursively divide each node into two child nodes $R_1(j,s)=\{x|x^{(j)}\leq s\}$ and $R_2(j,s)=\{x|x^{(j)}>s\}$ according to a feature $j$ and a threshold $s$. To minimize the MSE of the final tree, we choose $(j,s)$ to minimize the MSE of each division, 
\begin{align}
    \mathrm{MSE}_{(j,s)}=\frac{1}{n} \left[ \sum_{\bm x_i \in R_1(j, s)} (y_i - c_1)^2 + \sum_{\bm x_i \in R_2(j, s)} (y_i - c_2)^2 \right],
    \label{eq:bestsplit}
\end{align}
%More specifically, starting from the root node, the following operations are performed recursively on each node. For each candidate split $(j, s)$ consisting of a feature $j$ and threshold $s$ that partitions the parent node data into $R_1(j,s)=\{x|x^{(j)}\leq s\}$ and $R_2(j,s)=\{x|x^{(j)}>s\}$,
%we choose $j$ and $s$ by minimizing the MSE from the division, 
where $c_1$ and $c_2$ correspond to the averages of the labels $y_i$ in $R_1$ and $R_2$. The division continues till some stopping criteria regarding the depth of the tree or the size of the leaf node are satisfied, which we specify in section~\ref{sec:training}. %Perform the above operations on each node until the stopping conditions are satisfied. A complete decision tree form.

%To find the best split $ \theta^* = (j^*, s^*)$ in each node, first fix feature $j$ and then select the best division $s$ under that feature according to equation~\ref{eq:bestsplit}. Doing this for each feature, we can get the $m$ best divisions as there are $m$ features. 
%Among these $m$ combinations of $(j,s)$, the one with the minimum value is the globally optimal $(j^*, s^*)$.
Once a tree is constructed, it is straightforward to make predictions with it. A new input observation $x_{\rm new}$ can be inserted into a leaf node through a tree walk, and the corresponding prediction is found as the average $y$ of the training data in the leaf.
% \begin{align*}
%  y_{new} = \frac{1}{m}\sum_{x_i \in L}y_i 
% \end{align*}
% where $L$ is the set of elements $x_i$ in leaf nodes and $m$ is the number of data in the leaf.

\subsection{Random forest}
A decision tree can easily overfit the data. For example, when a leaf node contains only one observation, any noise in the observation will be inherited by the model prediction. To overcome this problem, a random forest works by combining the predictions of many trees each constructed from a bootstrap realization of the original data. 
%Thus the input data set of each tree is different, and that ensures the diversity of the trees in the forest and improves the generalization ability.

For each tree in the forest, when selecting splitting features on a node, a further pooling step is added to restrict the selection to a random subset of the original feature set. This randomness further enhances the generalization ability of the model.
The final prediction is obtained by combining (averaging in the case of regression) the predictions of all trees. 
% \begin{align*}
%  y_{\rm final} = \frac{1}{M}\sum_{m=1}^{M}y_m 
% \end{align*}
% where $M$ is the number of trees in the forest, and $y_m$ is the prediction from tree $m$.
\subsection{Feature importance and Exhaustive Feature Selection}
A random forest can not only serve as a predictive model that fits the data. It can also output an importance score for each feature quantifying its relative contribution in the prediction,
%In many cases, the goal is not only to build a predictive model but also to determine which predictor variables are most important in making these predictions. Just as we are interested in which satellite galaxies are most important for halo mass estimation in this work. Feature importance is used as an index to measure the contribution of features to prediction, 
which is of great significance in feature selection and helps to understand the underlying model construction process. In the \textsc{scikit-learn}~\citep[]{scikit-learn} package, RF feature importance ranking is based on the Mean Decrease Impurity (MDI), which quantifies the average reduction in MSE contributed by the tree divisions in each feature. We provide the detailed definition of the MDI importance in Appendix~\ref{app:MDI}.

%For a Random Forest with $M$ trees, the importance of a feature $j$ is obtained by further averaging the importance of $j$ in each of the trees.% single tree $\phi_m$:
% \begin{align*}
% Imp(X_j)=\frac{1}{M}\sum_{m=1}^{M}\sum_{t\in \phi_m}p(t)\Delta I(s_t, t)
% \end{align*}

%\subsubsection{Exhaustive feature selection}
Despite that the feature importance in random forests based on MDI is widely used for feature selection, it has been shown in the literature that such importances may produce misleading results~\citep[]{Strobl2007_RFbias,Louppe14,Scornet2020_RFbias}. For completely independent variables and in absence of variable interactions, MDI provides a variance decomposition of the output. However, for partially redundant variables that carry similar information, which almost always happen in practice, the one with slightly more information may always stand out in the feature selection process of node splitting, leaving little MDI to the others.
%If two features are strongly correlated, then the importance of one of them may be masked, resulting in a bias in importance ranking. 
For this reason, we can not completely rely on the importance ranking given by the random forest. Hence we take the strategy of still using the random forest as the regression model while combining it with an exhaustive method for feature selection.

Specifically, we try all possible feature combinations and train one model using each combination. The performances of the models are then compared to select the best combination of features for each number of features. The best feature combinations at each step are selected according to the $\rm R^2$ score, which is used to evaluate the performance of regression models.
The $\rm R^2$ score is defined as 
\begin{align}
  \mathrm{R^2}= 1-\frac{\sum_{i}(y_i-\hat{y}_i)^2}{ \sum_{i}(y_i-\bar{y})^2},
  \label{eq:R2}
\end{align}
where $y_i$ is the true target variable with $\bar{y}$ being its mean value in the sample, and $\hat{y}_i$ is the predicted value for observation $i$.
This approach enables us to identify the most important feature combinations without having to worry about feature correlations. The evolution of the performance with the addition of features can also be used to understand the unique contributions of features to the model improvement.
%Sequential forward selection is a type of heuristic search algorithm, which is actually a simple greedy algorithm. The basic idea is that the feature subset $X$ starts as the empty set and picks one feature $x_j$ each time to add to the feature subset $X$, which means that one feature is selected at a time to add to the feature set that enables the evaluation function to take the optimal value.
\subsection{Tuning the hyperparameters}\label{sec:training}
To achieve the best performance of a machine learning model, it is crucial to tune the hyperparameters of the model. For RF in \textsc{scikit-learn}, there are several major hyperparameters to be tuned: \texttt{n\_estimators}, \texttt{max\_depth}, \texttt{min\_samples\_leaf} and \texttt{min\_samples\_split}.
We start the tuning process from  \texttt{n\_estimators}, and obtain the number of trees in a range that makes the model perform best. The remaining parameters are further tuned one by one after fixing previous parameters to their optimal values. The final set of adopted hyperparameters are presented in table (\ref{Tab:Tuning results}). 
\begin{table*}[ht]
\begin{center}
\caption{The optimal RF hyperparameters in our model. From left to right: number of trees within the forest, maximum growth depth of decision tree, minimum number of samples of leaves and minimum number of samples of branch nodes to split.}
\label{Tab:Tuning results}
\begin{tabular}[c]{c*{4}{c}}
 \hline
parameters & \texttt{n\_estimators} & \texttt{max\_depth} & \texttt{min\_samples\_leaf} & \texttt{min\_samples\_split}\\
 \hline
values & 190 &11& 10&10 \\
\hline
\end{tabular}
\end{center}
\end{table*}

\section{Results}
\label{sec:result}
%In the following subsections, we will discuss the main results obtained so far. 
%In this section we present the results from our RF model, including the model performance, the classical feature importance and feature selection through the exhaustive method.

%In Section~\ref{sec:performance}, we present the performance of our model in the form of predicted-true value relation and residual analysis. Then, in the following subsections, we show the importance ranking results of the random forest, and an exploratory analysis based on this result. Subsequently, in Section ~\ref{sec:sfs rank}, the consequences offered by the Sequential Forward Selection are demonstrated, and a few features are focused on for analysis.
\subsection{Model Training and Performance}
\label{sec:performance}
\begin{figure}[htbp!]
  \centering
  \includegraphics[width=\columnwidth]{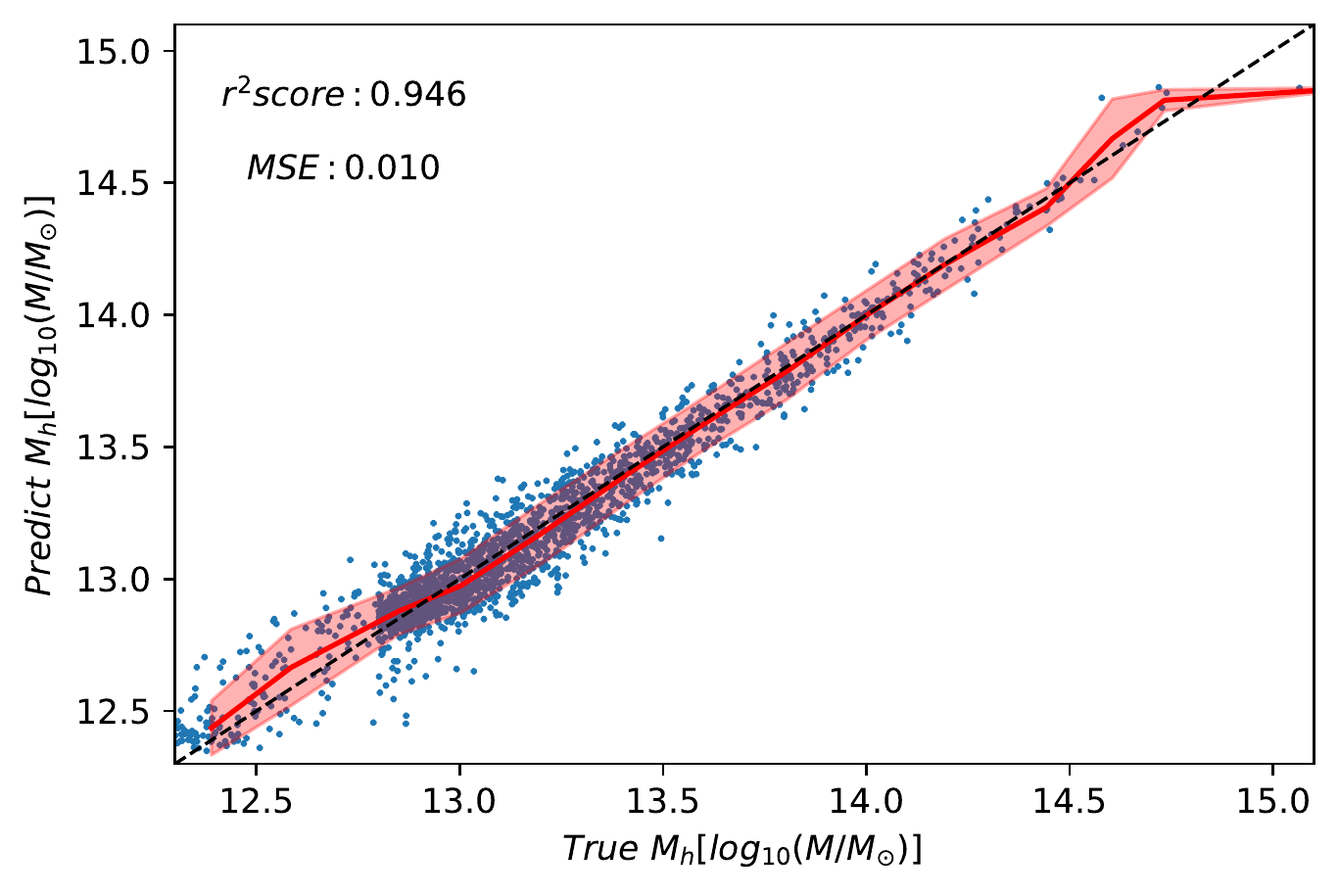}
  \caption{Relationship between the predicted halo mass and true halo mass values. The diagonal dashed line is the line represented $y=x$, and red line shows the median relation between true value and predict value. Red shaded region represents the $1\sigma$ percentile. The data points are spread uniformly and concentrated on both sides of the line, indicating that the difference between the predicted and true values is quite small and that random forest makes a good prediction of the halo mass. The fitted $\rm R^2$ score is 0.946 and the Mean Square Error (MSE) value is 0.01.}
 \label{fig:predicted vs. true}
\end{figure}

Our fiducial model is the random forest model that adopts the hyperparameters given in Table (\ref{Tab:Tuning results}), taking the logarithmic stellar masses of the top seven galaxies in the mass ranking as the input feature variables, and the target variable to be predicted is the logarithmic dark matter halo mass. Cross-validation was employed to evaluate the model performance, dividing the dataset into a training set and a test set, which are used for training and testing respectively. %Training the model on the training data and testing the performance on the test data, the model could achieve an fairly high R2 score\jx{what is R2, =MSE?} of 0.953, indicating that the random forest regression is working well.
Figure \ref{fig:predicted vs. true} shows the relation between the predicted and true halo masses in the test set of our model. Overall, the model can unbiasedly predict the true halo mass accross the entire mass range, with a fairly small total MSE of $0.01$ and a $\rm R^2$ score of $0.946$. The deviation of the few data points at the highest mass end is due to the limited number of haloes in this mass range that get allocated into a single leaf node. %As can be seen from the figure, RF can give a precise halo mass estimation.

Besides the fiducial model, we further train additional models using subsets of the available features as input. 

\begin{figure}[htbp!]
  \centering
  \includegraphics[width=0.9\columnwidth]{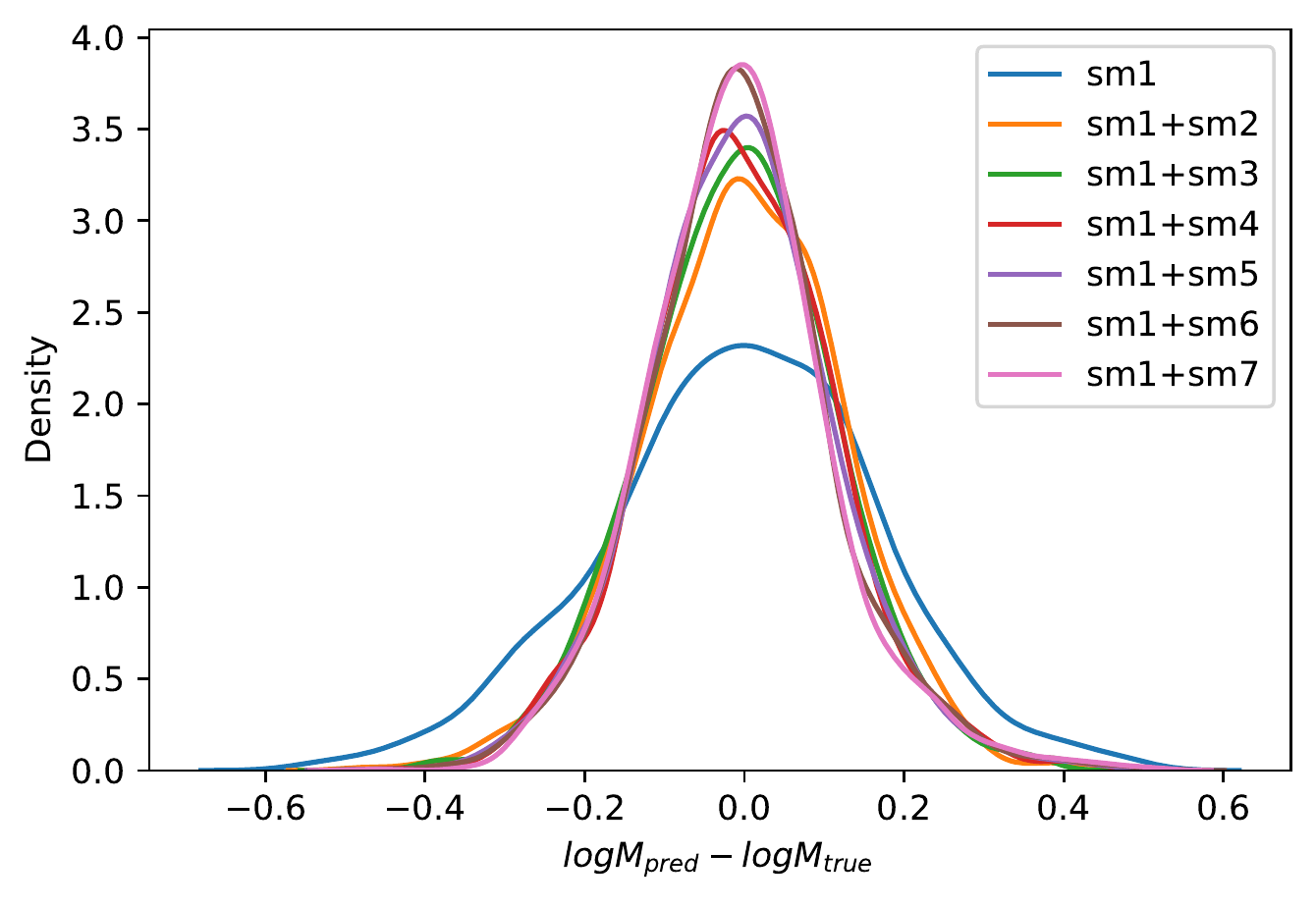}
  \caption{Probability density functions of the residuals from models trained with different combinations of the central (sm1) and one satellite (sm2-7) respectively. The inclusion of any of the satellite galaxies results in a tighter distribution in the residual. However, no single satellite significantly outperforms the others in the improvement.} 
  \label{fig:residual}
\end{figure}

In Fig.~\ref{fig:residual}, we compare the residual distributions for models involving the central and another satellite in the halo mass prediction.
%Furthermore, we carry out an analysis of the residual distribution of the models constructed by different combinations of features (Figure \ref{fig:residual}). To check if the satellites could improve the precision of the prediction, we compare the probability density functions of the predicted residuals using only the central galaxy and using the combination of the central galaxy plus a satellite galaxy to estimate the halo mass separately (Figure \ref{fig:residual}). 
As can be observed, the inclusion of a satellite galaxy leads to a more accurate prediction compared to that using only the central galaxy, indicating that satellite galaxies can indeed provide additional information for the halo mass estimation. However, the residual distributions involving different satellites are all very close to each other, suggesting that there is not an outstanding satellite that improves the prediction much more than the others. We will come back to this conclusion later. 

In the following we explore the roles played by different galaxies in more detail using feature importance and exhaustive feature selection. %it is not too easy to directly compare from this figure alone which satellite galaxies are present with the most significant contribution.

\subsection{Importance Ranking}
\label{sec:RF result}
\begin{figure}[htbp!]
  \centering
  \includegraphics[width=\columnwidth]{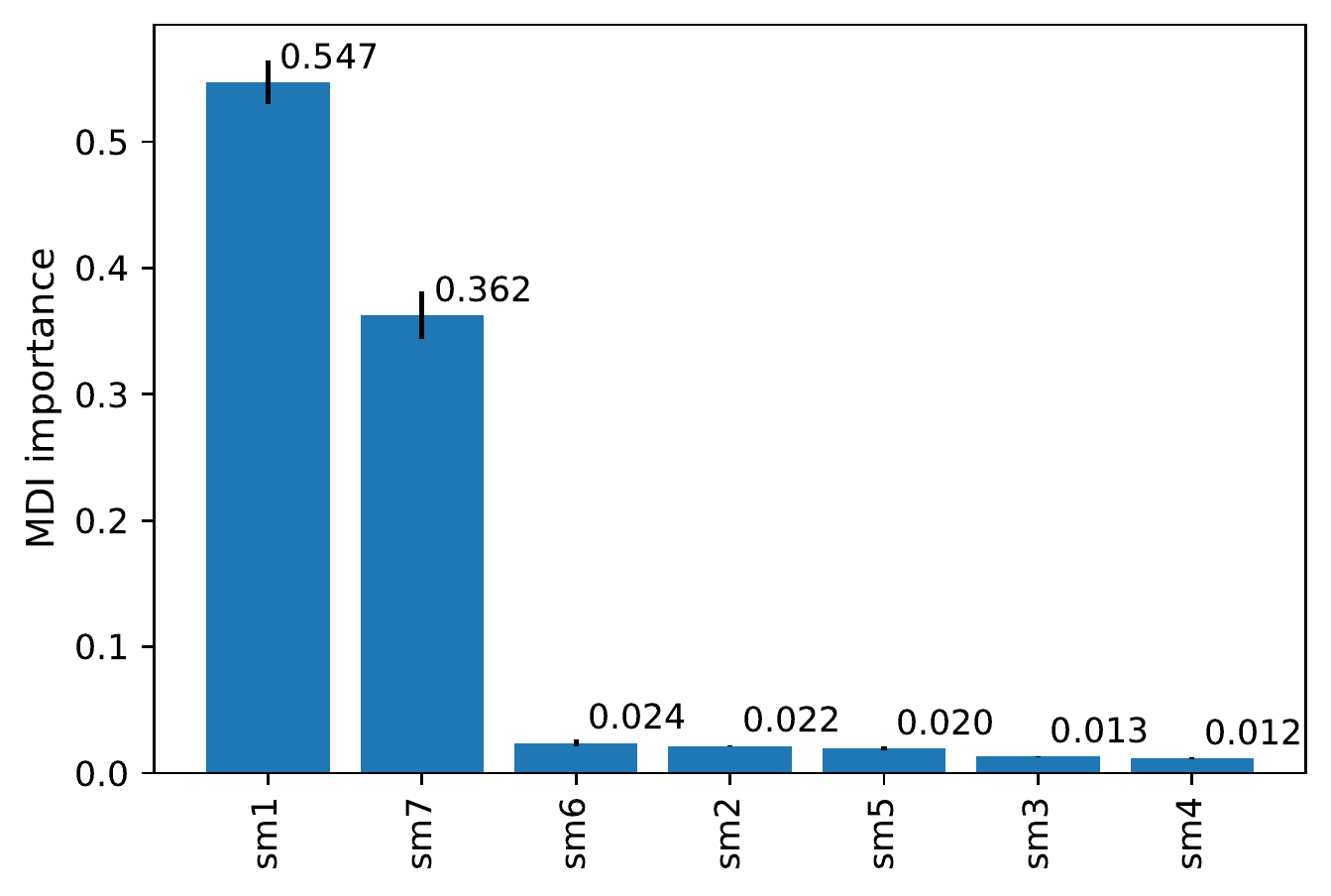}
  \hspace{1cm}
  \caption{MDI based feature importances given by the random forest. Histogram height represents the relative importance and the error bar is the standard deviation of the importance from individual trees in the forest.}
  \label{fig:importance}
\end{figure}

The MDI based importance ranking given by RF is presented in Figure~\ref{fig:importance}. The error bars are the standard deviations of the importances from individual trees in the forest. As expected, the most important feature is the stellar mass of the central galaxy. The second important feature is the stellar mass of the 7th massive galaxy, which is also the least massive galaxy in the data we used.

As mentioned above, Exhaustive Feature Selection (EFS) was introduced considering that the default feature importance ranking provided by random forest might carry some bias. We train our model using all possible feature combinations, and list the top four best scoring combinations for each number of features in Table~\ref{tab:results of efs}. A 5-fold cross validation is adopted in this process. Specifically, we split the dataset into 5 equal subsets (a.k.a., folds) with each subset used once as validation while the 4 remaining folds forming the training set. The error of  $\rm R^2$ score is calculated as
\begin{align}
  \sigma_{R^2}= \sqrt{\frac{\sum_{i}(s_i-\bar{s})^2}{k(k-1)}},
  \label{eq:R2 error}
\end{align}
where $s_i$ is the $\rm R^2$ score of fold $i$ with $\bar{s}$ being the mean score, and $k$ is the fold numbers.

\begin{table*}
\renewcommand\arraystretch{1.5}
\begin{center}
\caption{The top four scoring feature combinations from exhaustive feature selection. The numbers in the brackets specify the stellar mass ranks (1 for central and 2-7 for satellites) of the constituting galaxies. The scores are the R$^2$ scores of the corresponding model.}
\resizebox{\textwidth}{!}{
\begin{threeparttable}
\label{tab:results of efs}
\begin{tabular}[c]{cllllllll}
 \hline
 \multirow{2}{*}{Feature number}&
 \multicolumn{2}{c}{rank1}&\multicolumn{2}{c}{rank2}&\multicolumn{2}{c}{rank3}&\multicolumn{2}{c}{rank4}\cr
 \cmidrule(lr){2-3} \cmidrule(lr){4-5} \cmidrule(lr){6-7} \cmidrule(lr){8-9} 
    &Feature&Score&Feature&Score&Feature&Score&Feature&Score\cr
\hline
1&[1]&0.8395$\pm$0.00283&[7]&0.8263$\pm$0.00219&[6]&0.8144$\pm$0.00337&[5]&0.7913$\pm$0.00572\cr
2&[14]&0.9246$\pm$0.00057&[13]&0.9244$\pm$0.00136&[15]&0.9244$\pm$0.00136&[16]&0.9239$\pm$0.00119\cr
3&[127]&0.9440$\pm$0.00064&[126]&0.9423$\pm$0.0076&[125]&0.9407$\pm$0.000104&[137]&0.9402$\pm$0.00012\cr
4&[1237]&0.9466$\pm$0.0008&[1247]&0.9463$\pm$0.00075&[1257]&0.9457$\pm$0.00082&[1267]&0.9447$\pm$0.00073\cr
5&[12347]&0.9471$\pm$0.00083&[12357]&0.9471$\pm$0.00093&[12367]&0.9467$\pm$0.00088&[12457]&0.9464$\pm$0.00081\cr
6&[123457]&0.9472$\pm$0.00086&[123467]&0.9471$\pm$0.00084&[123567]&0.9470$\pm$0.00095&[124567]&0.9464$\pm$0.00082\cr
7&[1234567]&0.9471$\pm$0.00088\cr
\hline
\end{tabular}
\end{threeparttable}}
\end{center}
\end{table*}

\begin{figure}[htbp!]
    \centering
    \includegraphics[width=\columnwidth]{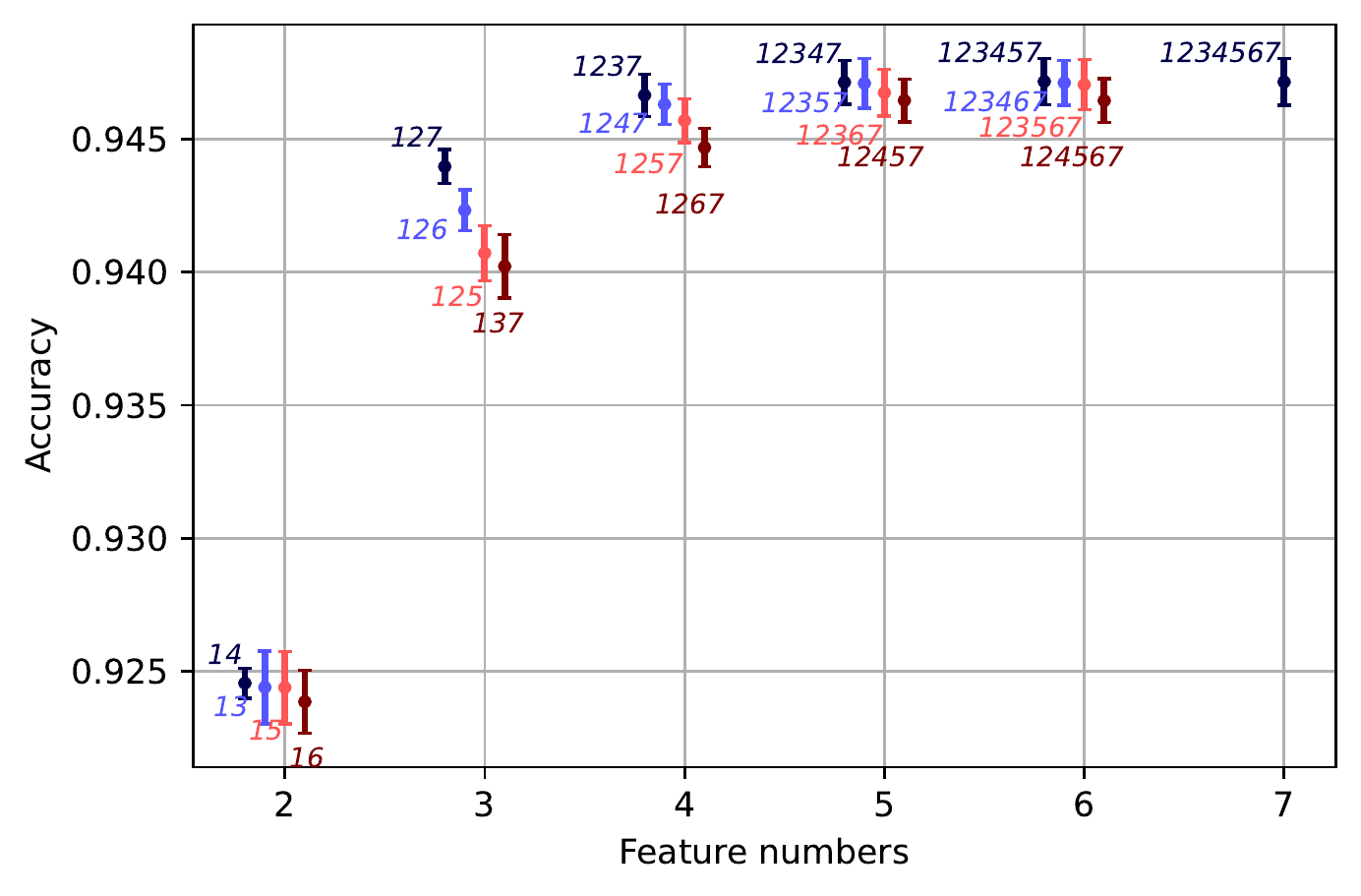}
    \caption{Performances of the top four scoring feature combinations for each number of features. The data points are the R$^2$ scores of the 5-fold cross-validation and the errorbars are their standard deviations. Different colours represent different rankings, slightly offset horizontally for better visibility.
    The detailed combination names are labelled next to each point. The scores of solo features are not plotted to reduce the dynamical range of the figure. }
    \label{fig:exhaustive all}
\end{figure}

In Fig~\ref{fig:exhaustive all}, we plot the performance of the models trained by different combinations versus the number of features. It is seen that the scores of models trained with two features improve compared to model trained only by the central galaxy in Table~\ref{tab:results of efs}. It verifies once again that satellite galaxies have an extra contribution to the prediction of halo mass. However, by observing the case when only two features are available, we can see that the scores of the different combinations do not differ significantly and that no combination is outstanding. That is, the satellite galaxies of different orders alone play a similar role as complements to the central galaxy in the prediction of halo mass. 
For the case when only three features are input, the [127] (stellar masses of the first, second and seventh galaxies) combination gives the highest model score and is almost as high as the highest score attainable. This means instead of the whole population, we can use the information from only the first, second and seventh (here the least massive) galaxies as a high precision probe of halo mass.
Moreover, once the input feature number reaches 4, the improvement of the model goes less noticeable and even almost absent with further increase in feature number. This result indicates that information of only a few satellite galaxy members is sufficient to make high-precision predictions of the halo mass regardless of a complete galaxy population. 

\subsection{The roles of different galaxies}
In order to disentangle the role played by different ranked galaxies for the prediction of the halo mass in more detail, we choose the first, second, fourth and the seventh ranked galaxies to analyse.

\begin{figure*}[htb]
\centering
\includegraphics[width=1.8\columnwidth]{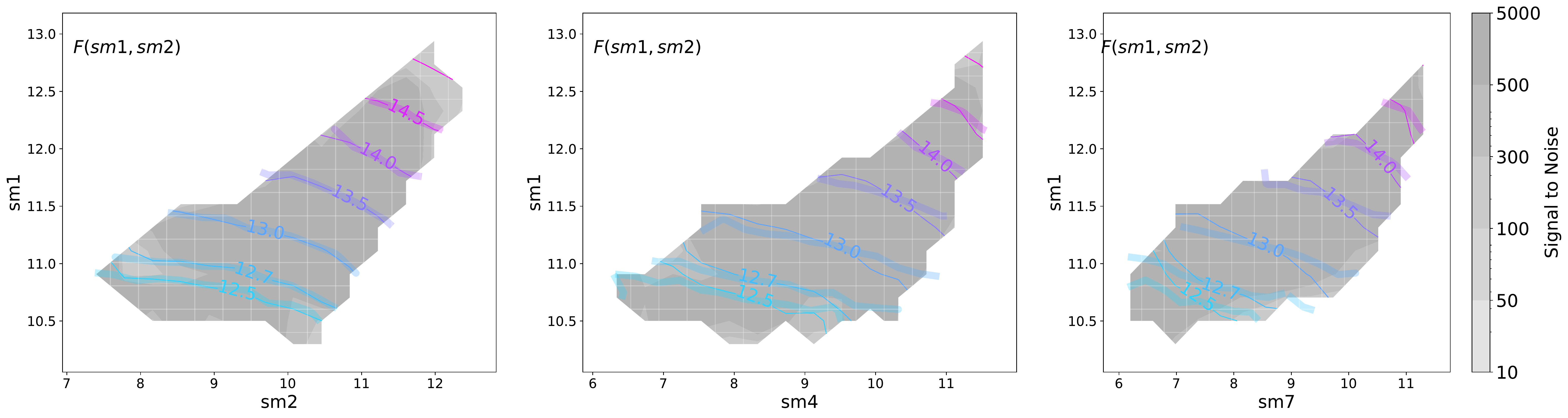}
 \hspace{1cm}    
\includegraphics[width=1.8\columnwidth]{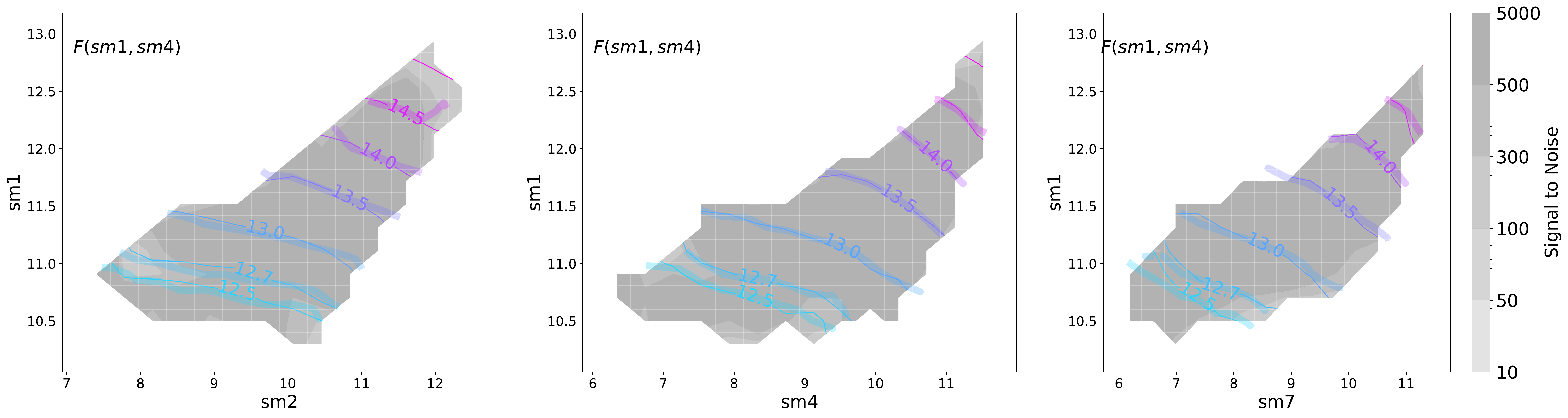}
\hspace{1cm}                 
\includegraphics[width=1.8\columnwidth]{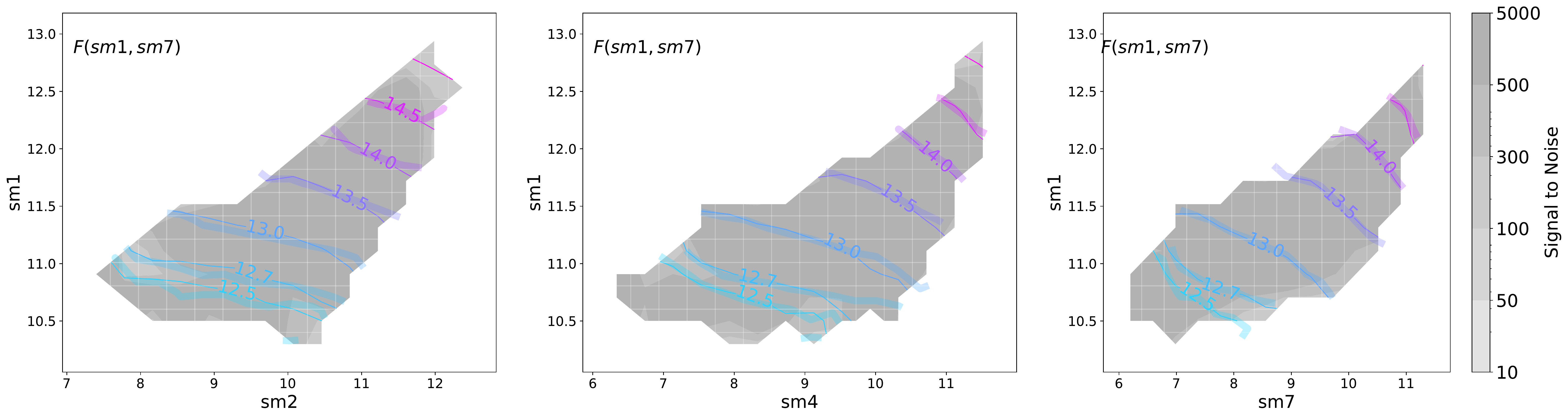}
\caption{Relationship between the  predicted halo masses from different models and stellar masses of different galaxies. The axes are the logarithmic stellar masses of the central (sm1), second (sm2), fourth (sm4) and seventh (sm7) most massive galaxies. 
%The grey background region reflects the distribution halos in each parameter plane. 
The grey filled contours show the signal-to-noise level of the data at each location, which reflects the number of halos within each bin. The thin coloured lines are the contours of the true halo masses, while the thick light lines are the contours of the predicted halo masses of the corresponding model labelled in each panel.}
\label{fig:2D biasmap}
\end{figure*}

We first consider the combinations of the central galaxy with one satellite galaxy, i.e., [12],[14] and [17], and train a model for each combination. Then we plot the predicted as well as true values of these models as functions of two stellar masses at a time in Figure~\ref{fig:2D biasmap}.
%The grey background region reflects the distribution of halos in the corresponding parameter plane, with the finer lines being the contours of the true halo masses while the lighter thicker lines being the contours of the predicted halo masses.
Overall, the contours that represent the halo masses are roughly perpendicular to the axes corresponding to the stellar mass of central galaxy, suggesting that the halo mass can be mostly determined by the stellar mass of the central galaxy. However, the slight inclination towards the x-axis indicates that it also depends on the satellite galaxies. 

Taking the top row of the figure as an example, as the model is trained by the stellar mass of central and the second galaxy, the predicted and true values match well in the sm1-sm2 plane (left panel), confirming that the model have sucessfully learned the mapping between halo mass and these two stellar masses. While for the rest right figures, obvious misalignment exists between the true and predict contours. This reflects the difference in the information provided by the different galaxies for predicting halo mass. It implies that although the inclusion of different satellite galaxies provides roughly equivalent improvement to the halo mass estimation with central galaxy alone, the supplementary information relative to the central galaxy that they carry is different. The deviation between the predicted and true masses is larger in the rightmost panel, indicating a larger information difference between (sm2, sm7) than that between (sm2, sm4). This could also explain why the combination [127] is the best when we use only three features. 

%\subsection{The role of the least massive galaxy}

\begin{figure*}[htb]
   \centering
   \includegraphics[width=\columnwidth]{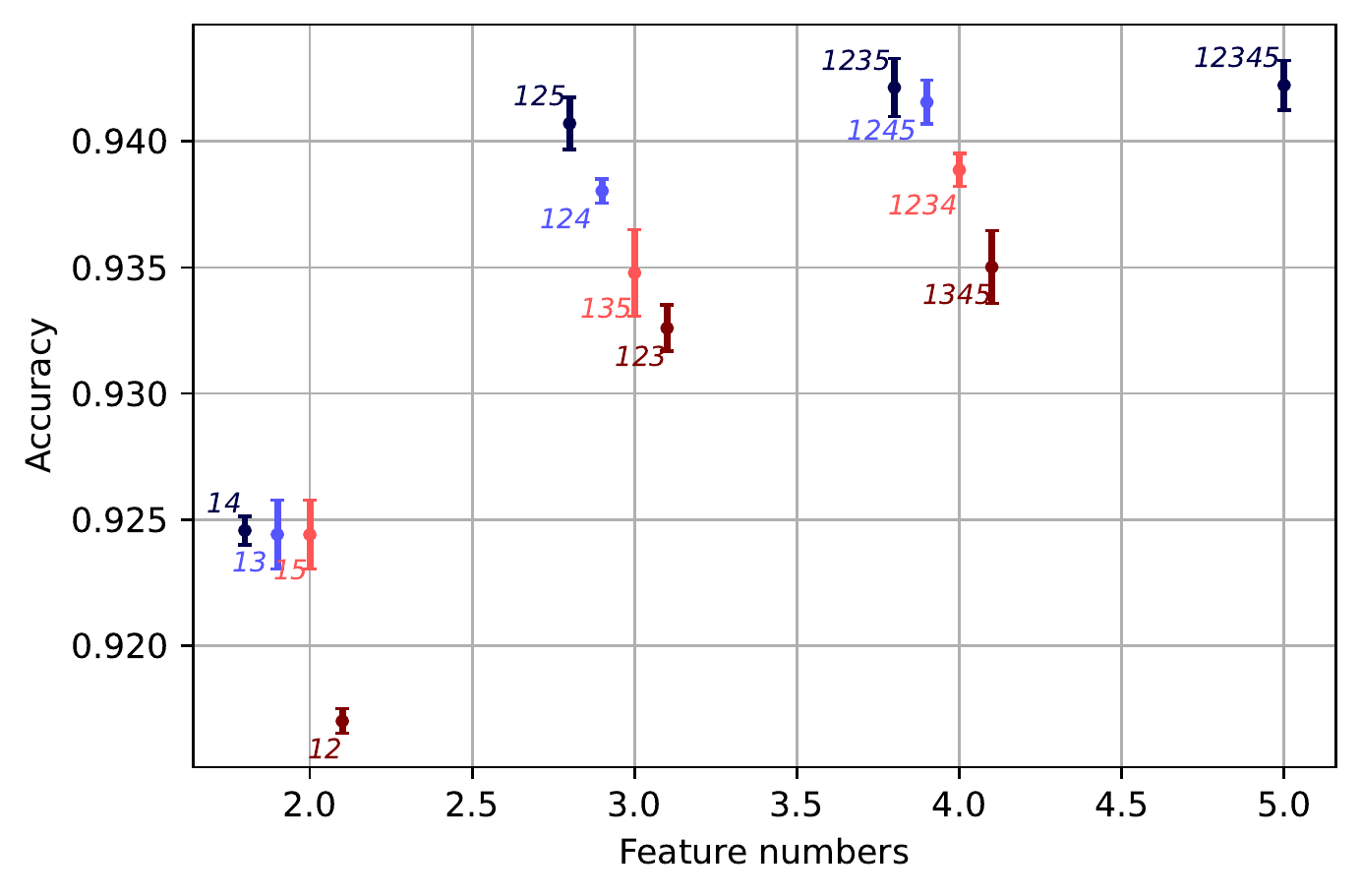}
  \includegraphics[width=\columnwidth]{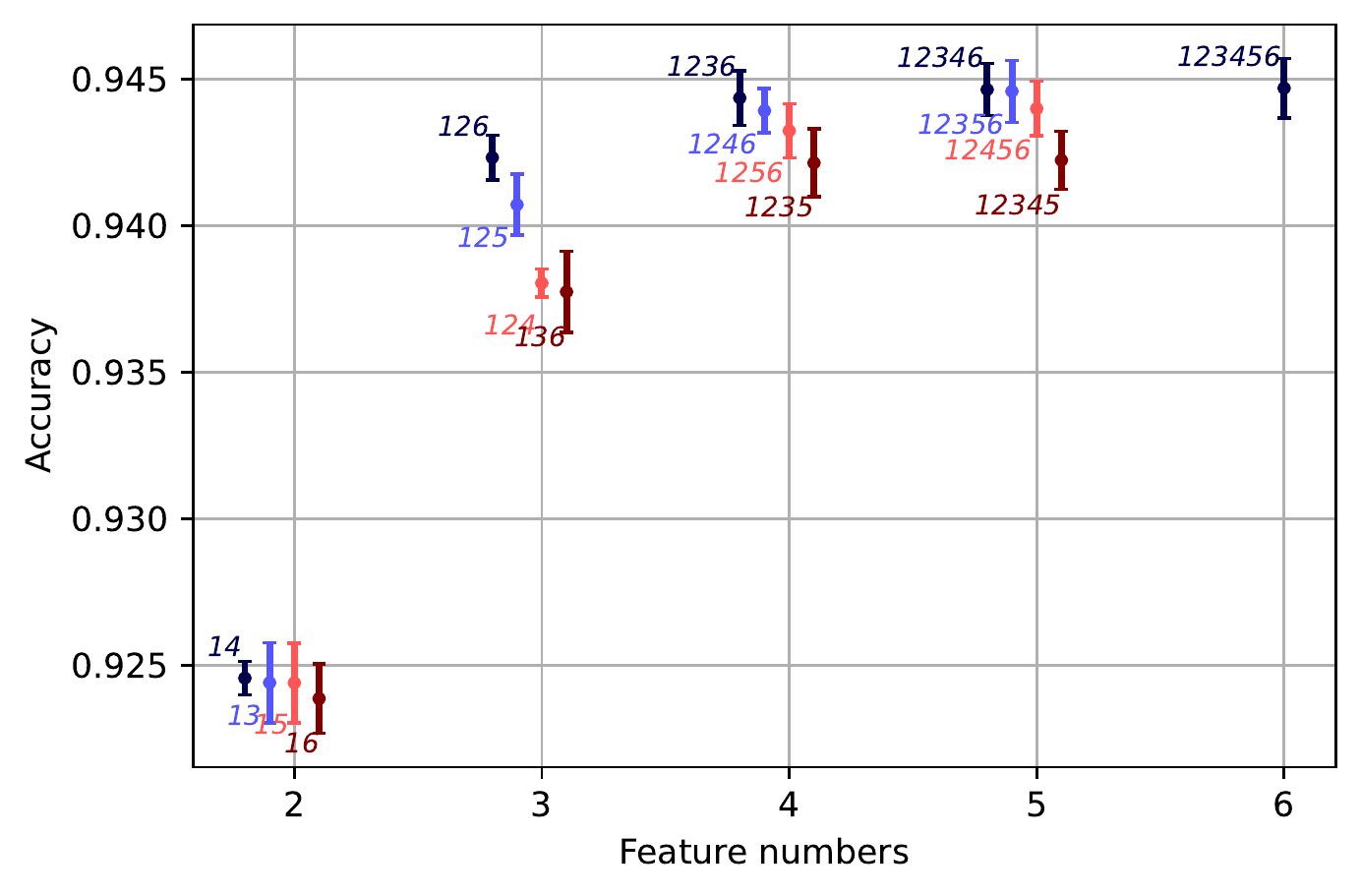}
   \caption{Similar to Figure~\ref{fig:exhaustive all}, but using halo samples containing at least 5 (left) and 6 (right) galaxy members respectively.}
   \label{fig:5&6 features test}
 \end{figure*}

It is interesting to note that the galaxies [127] also appear in the top combinations involving larger numbers of features in Table~\ref{tab:results of efs}.
The substantial extra information carried by the 7th satellite relative to the 2nd may be because it is the least massive satellite in our halos. In other words, the largest differences exist between the satellite galaxies with the largest ranking separation.  To further verify this interpretation, we perform the same analysis on complete samples containing 5 and 6 galaxy members respectively. The results are consistent: the best combination of three galaxies is always the first, second and smallest galaxies, as shown in Figure~\ref{fig:5&6 features test}.

\subsection{Mass range independence}
\begin{figure*}[htb]
  \centering
  \includegraphics[width=\columnwidth]{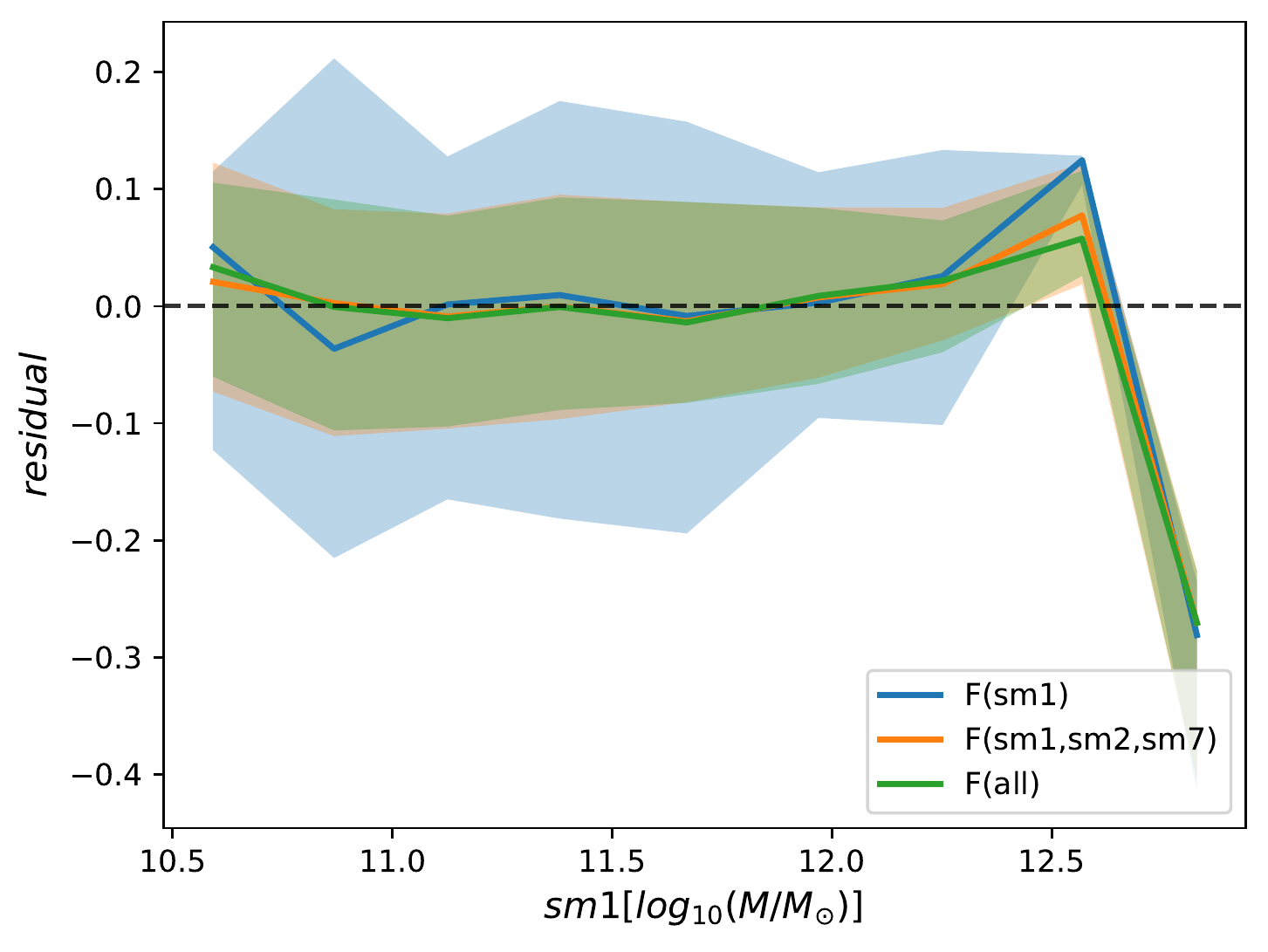}
  \includegraphics[width=\columnwidth]{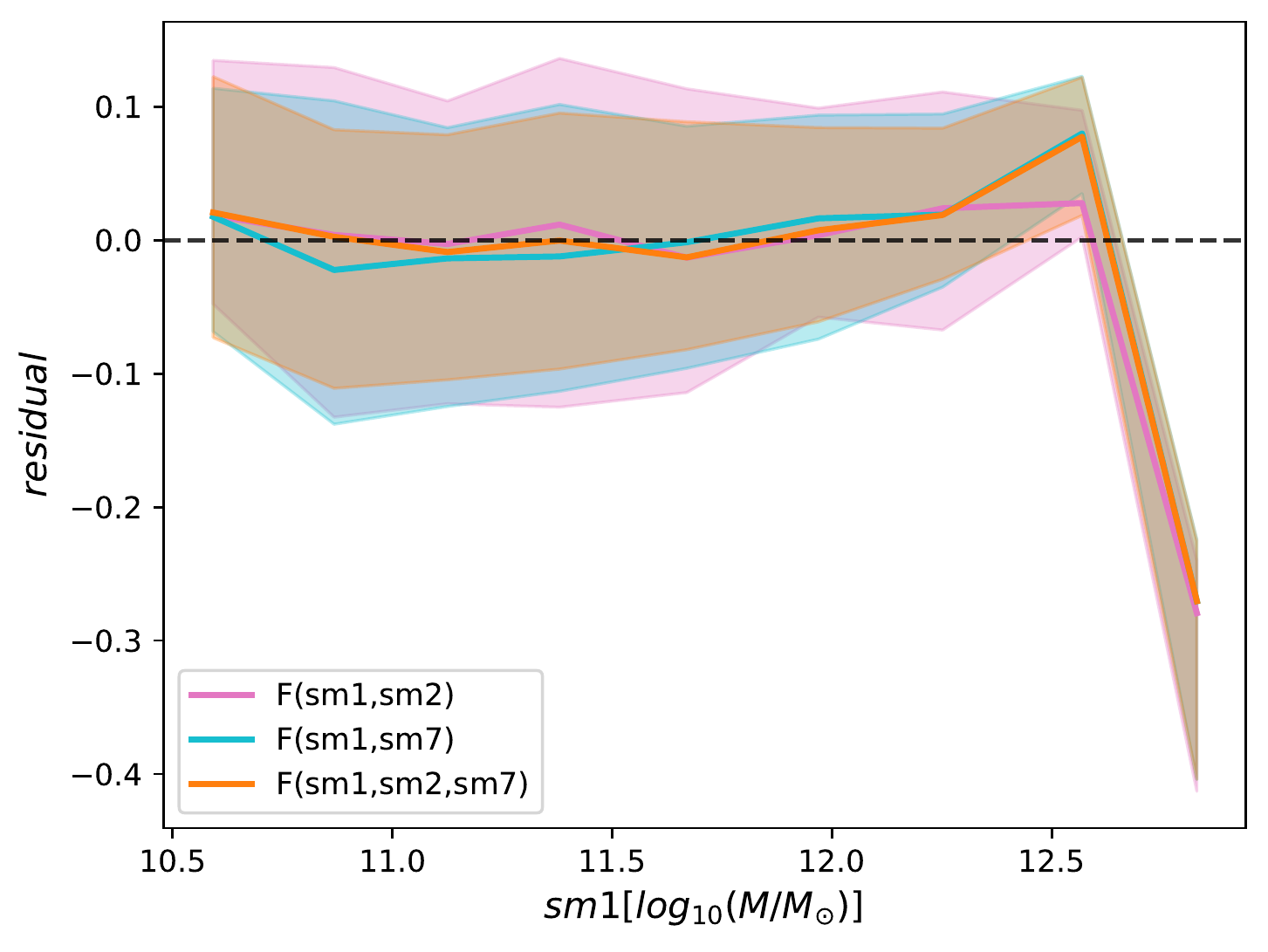}
  \caption{Residual distribution of different models versus the stellar mass of the central galaxy. Left panel: Distribution of residuals as a function of the stellar mass of the central galaxy for models trained with the stellar mass of central galaxy ($sm1$), central galaxy and the second, seventh galaxy ($sm1,sm2,sm7$), and all galaxies, respectively. Right panel: Distribution of residuals as a function of the stellar mass of the central galaxy separately for models trained with sm1\& sm2, sm1\&sm7, sm1\&sm2\&sm7. 
  The solid lines show the median relation of the residual and the stellar mass of central galaxy, and the shaded area shows the 1-$\sigma$ confidence region of residual values.The horizontal dark dashed line is for residuals equal to 0.}
  \label{fig:residual_mass dependence}
\end{figure*}

To guarantee that our results are not dependent on the mass range, we examine the residuals of different models at different central galaxy masses in Figure~\ref{fig:residual_mass dependence}.  
The residuals are concentrated around 0 in the full mass range, indicating that the models are unbiased over the whole mass range. Note the large deviation at the highest masses is due to the rarity of halos there. In addition, for the same model, the scatter is also similar over various masses. This suggests that our results have no dependence on mass and are valid throughout the mass range. Comparing the dispersions in the residuals, the previous conclusions can also be seen, that the inclusion of satellite galaxies helps to improve the accuracy of the halo mass estimation, and that the precision of the model constructed using the three best combinations is already comparable to that using all the features.

\section{Discussion: understanding the gaps with the conditional galaxy distribution}
\label{sec:CMF}
%So far, we have already drawn a conclusion that taking three galaxies as input features, the [127] combination is the best. As well as the studies on the magnitude or stellar mass gap~\citep[e.g.][]{Deason2013,Lu2015,GoldenMarx2018,Hearin2013} and Bradshaw's~\citep{Bradshaw2020} "cen+N" model, these informative galaxies are associated with a rank according to some criterion (magnitude or stellar mass). 

The magnitude or stellar mass gaps, and the galaxy combinations studied here are all constructed based on the ranks of galaxies. Such ranks and their corresponding sizes naturally appear as function values and random variables in the cumulative mass or luminosity functions. Such a connection have been exploited before to derive the distribution of magnitude gaps as well as that of the BCGs by drawing from the global or conditional luminosity functions~\citep{More2012,Paranjape12,Hearin2013,Shen2014,Paul17}. 

Unlike previous studies, our machine learning results allow us to explore this connection in a reverse manner, to directly identify where and how much information on halo mass is stored in the cumulative galaxy distribution. In this context, galaxies with different mass ranks control different segments of the cumulative stellar mass function (CSMF). Those with ranks 1 and 2 control the shape of the curve at the massive end, while those with rank 7 control the shape of the curve at the more distant end, i.e. the low mass end. The connection of the gap or rank statistics to halo mass can then be understood as the variation in the relevant segments of the conditional cumulative stellar mass function (CCSMF) with halo mass, $\phi(>M_\star|M_h)$. The finite number of informative features then reflects the limited number of distinct mass-dependent features in the CCSMF, or the universality of CCSMF subject to a few mass-dependent parameters. 

\begin{figure*}[htbp!]
\centering
\includegraphics[width=1.8\columnwidth]{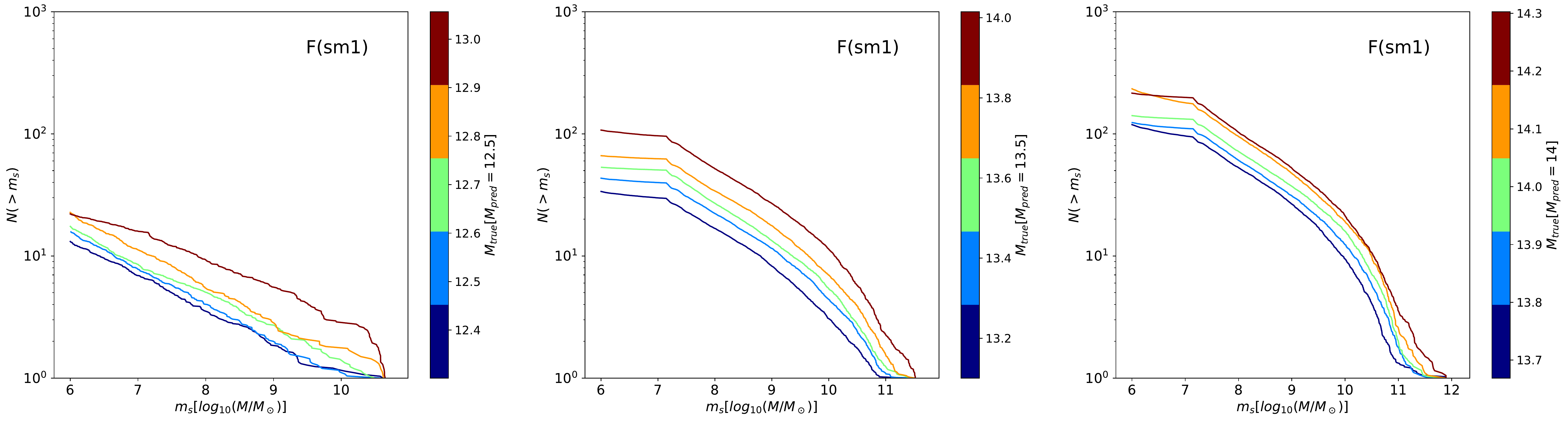}
 \hspace{1cm}    
\includegraphics[width=1.8\columnwidth]{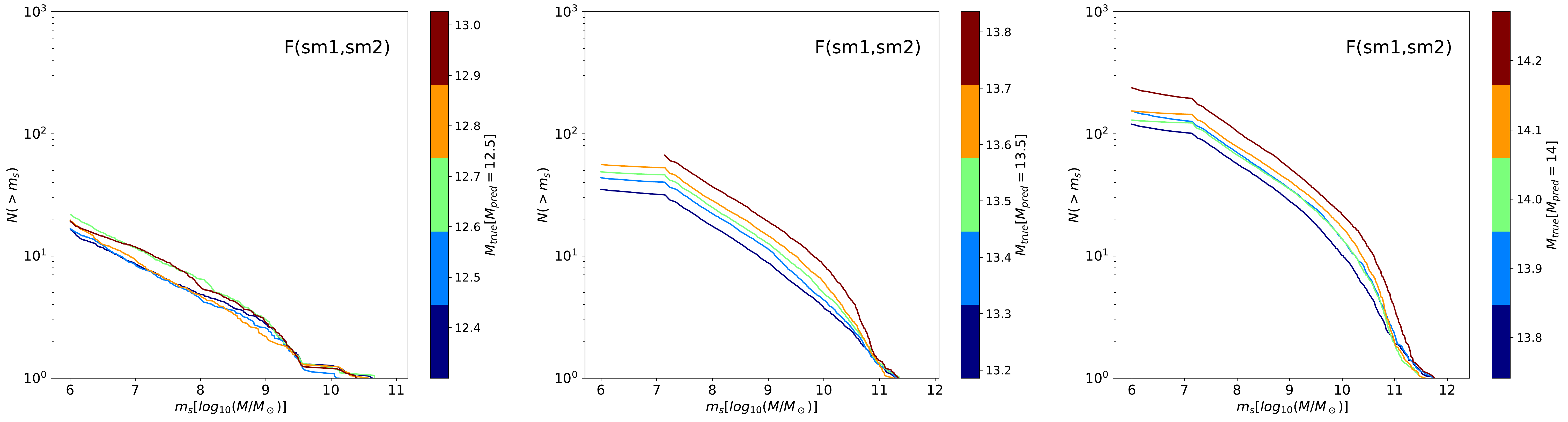}
\hspace{1cm}                 
\includegraphics[width=1.8\columnwidth]{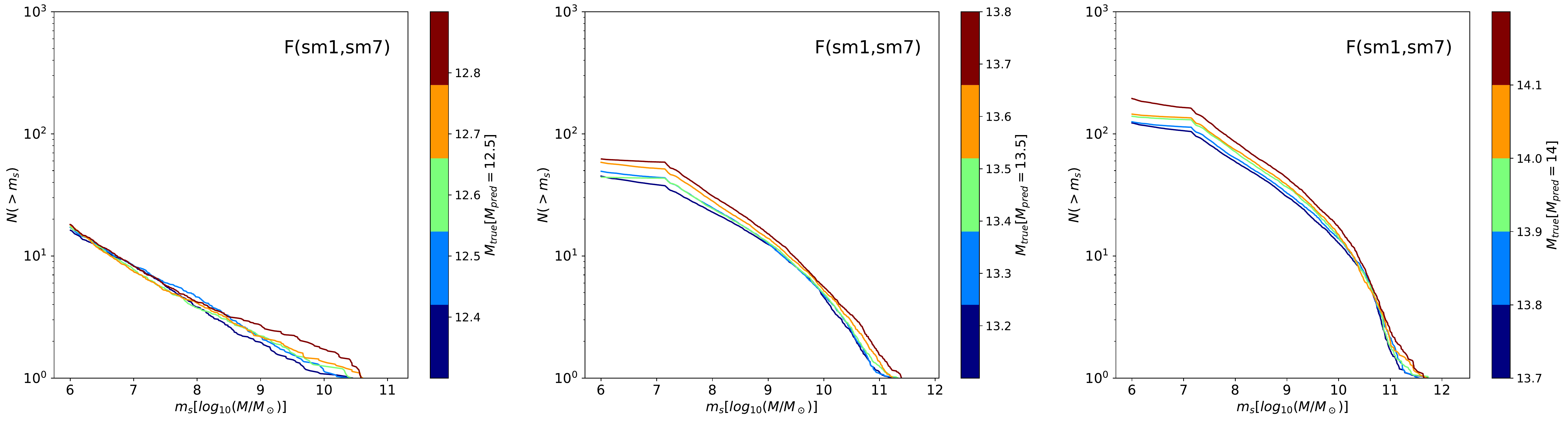}
\hspace{1cm}                 
\includegraphics[width=1.8\columnwidth]{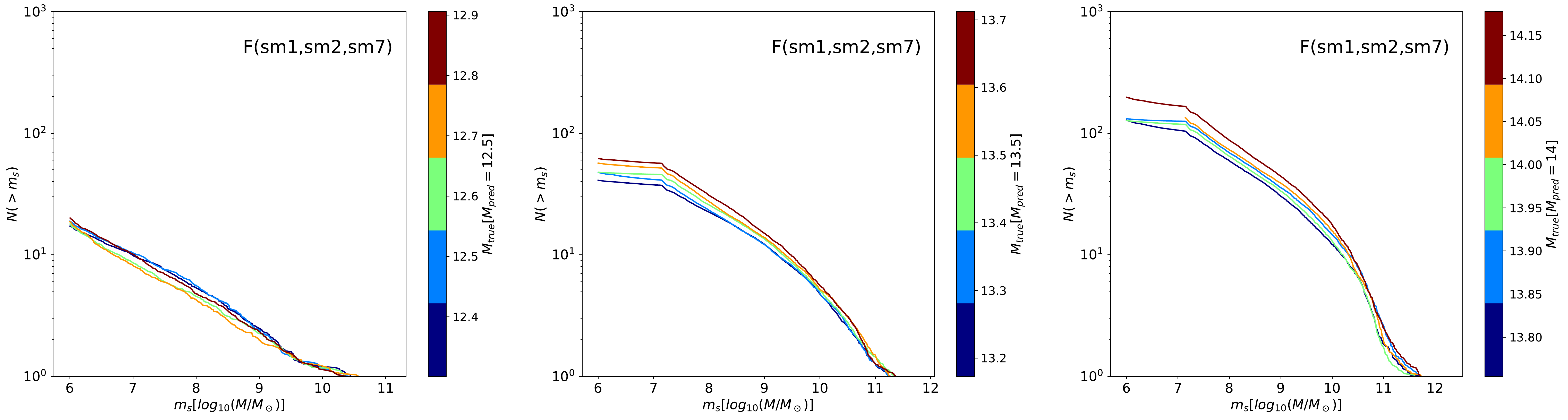}
\caption{The cumulative stellar mass function (CSMF) for halos with the same predicted mass but different true halo mass. From top to bottom, models are trained with galaxies [1], [12], [17] and [127] as input features respectively. From left to right, the predicted value of halo mass are binned around $10^{12.5}M_{\odot}$, $10^{13.5}M_{\odot}$ and $10^{14}M_{\odot}$ respectively, with a bin width of $0.5$ dex. Each coloured curve represents the CSMF for a given true halo mass as labelled in the colour bars.}
\label{fig:CSMF}
\end{figure*}

To verify this conjecture, we plot the CCSMF for halos with the same predicted values but different true halo mass values in Figure~\ref{fig:CSMF}. For a given model, fixing the prediction is equivalent to fixing the values of the input features and the corresponding segments in the CCSMF. The remaining differences in the CCSMF curves for different true halo masses then reflect the contributions of features outside the combination used.%, and thus allow us to examine our assumptions. 
We test the CCSMF using the RF models constructed respectively with combinations [1], [12], [17] and [127], and show the results for three representative mass ranges centered at $\log (M/M_\odot)=12.5$, 13.5 and 14 in predicted mass.% with a bin width of 0.05 dex. For samples in each of the same predicted value bins, their true halo masses were quintupled and the corresponding CSMFs for the different quintiles were then compared, as shown in Figure\ref{fig:CSMF}.

%In Figure\ref{fig:CSMF}, the color bar scales from blue to red with respect to the lowest to highest quintiles of halo mass. 
It can be seen from this plot that for the model trained with only the central galaxies (first row), the CSMF curves for different true halo masses are noticeably different at fixed prediction, and only converge at the most massive end where the central mass is fixed. In the second row, adding the second rank galaxies to the model, the curves show a further strong tendency to bunch up at the massive end, but still with a clean separation in the low mass region. Correspondingly, in the third row, adding the seventh rank galaxies to the central galaxies, one of the focal points of the curves shift to the relatively lower mass end at $N=7$, and yet some discrepancies of the curves can be seen between and beyond the two focal points. Finally adding both the second and seventh galaxies as supplements to the central galaxy (last row), it is seen that the CSMF curves already approach complete overlap in the region of our concern ($N\leq 7$), exhausting the distinct features in the top 7 galaxies.

All these results are consistent with what we speculated earlier. When only the central galaxy is controlled, the apparent divergence between the curves implies that there is additional information beyond the central galaxy that we can utilise in the estimation of the halo mass. Further constraining both the central galaxy and the second or seventh galaxy, the previous divergence converges further at the corresponding massive and low mass end, and the halo mass is tightened further, suggesting that the inclusion of satellite galaxies improves the prediction and that the second and seventh galaxies contribute distinctively to their host halo. After restricting both the central galaxy and the second and seventh galaxies together, the separation between the curves in the area we considered almost disappears, indicating that the extraction of the information required for the prediction is almost maximised, which is consistent with the result in previous sections that the precision of the estimations is nearly saturated after the inclusion of the three best combined features. 

It is interesting to notice that for the middle and right columns where more satellites can be resolved in a halo, the CCSMF still diverges at the lowest mass end even in the [127] model. This means the low mass end distribution still carries extra information that can be used to further constrain the halo mass, in addition to that already explored in the top 7 members. It is also consistent with our previous finding that the least massive satellite in the sample can contribute significantly to the halo mass estimation, instead of galaxy 7 being special. This can be equivalently understood as the least massive satellite controls the overall amplitude of the faint end mass function or the richness of the halo, which is known to be tightly connected to halo mass.

\section{Conclusions}
\label{sec:summary}

In this work we have explored the connection of galaxy population to the host halo mass, to clarify the roles played by different galaxies on the halo mass estimation and to understand the information content of galaxy mass distribution on the halo mass. To this end we extract halos with at least 7 satellite galaxies from the IllustrisTNG simulation, and train a random forest algorithm to systematically assess the importances of different galaxy mass combinations in the prediction of halo mass. The results are further examined in the context of the conditional stellar mass function. 

Our findings and conclusions are summarised as follows.
\begin{itemize}
    \item When only one galaxy is used, we confirm that the central galaxy is the most informative single feature in estimating the halo mass.
    \item Compared with models that only use the central galaxy mass, the inclusion of satellite galaxy masses does improve the estimation of the halo mass, and the most informative binary features are always the central galaxy mass combined with another satellite mass.
    \item For the case of a combination of only two galaxies, the difference between the improvement of the model by adding any of the satellite galaxies to the central galaxy is not significant. This means there is not an outstanding satellite galaxy which contributes much more than the others to the halo mass estimation. In other words, we do not find an obviously ``optimal'' mass gap to be used in mass estimation. %Although there are some differences in the information available from different satellite galaxies, most of the information available for improving the estimate of the dark halo masses overlaps with each other. 
    \item For combinations of three galaxies, %although the difference between the performance of the central galaxy plus two satellite galaxies is not particularly apparent, 
    the best combination is always that of the central galaxy with the second and the least massive galaxies. This conclusion holds when examining the top 7, 6 or 5 galaxies, and may be generalised to a larger number of available galaxies. It suggests that the biggest and smallest satellite galaxies provide the greatest differential information. 
    
    \item For the seven member galaxies studied, the combination of a central galaxy and 2 or 3 satellite galaxies gives a near-optimal model performance, and continued addition of feature variables barely improves the model performance further. In other words, only a few galaxies are required to build a model with comparable accuracy to that using the whole member galaxy population.
    
    \item The different roles played by differently ranked galaxies can be directly mapped to the variation of different segments of the CCSMF with the halo mass. While the central galaxy controls the starting point of the CCSMF, the second massive galaxy controls the variation at the high mass end, and the least massive galaxy controls the amplitude or shape at the low mass end. Once these 3 galaxies are controlled, the CCSMF, that is, the full mass distribution of all the member galaxies, become largely determined in the studied mass range, with little extra variations that can inform about halo mass. However, we notice that the CSMF still contains extra variation at even lower masses beyond the 7th galaxy, which could be used to further constrain the halo mass.
    
\end{itemize}

The physical mechanism responsible for the information in the second and least massive galaxies might be that the former is related to recent as well as major merger events~\citep{Deason2013}, while the latter characterises the total mass accretion of the halo. Recent and major mergers can significantly influence the mass distribution around the halo, causing it to deviate from common galaxy-halo connections.  Moreover, the second massive galaxy and the smallest satellite galaxy in the satellite population have the greatest difference in the time of entry into the host halo, and therefore the greatest gap in the information they can provide. 

Our findings can provide insights into how to choose members to obtain the most information about the halo mass when the available galaxy population in the halo is limited. The direct visualisation of the CSMF dependence on halo mass also has implications on how to describe the CSMF, to maximize the information it carries about halo mass. It remains interesting to check whether current CCSMF or similarly conditional luminosity function models~\citep[e.g.,][]{Yang2003,Guo18} can fully capture these information. It is also straightforward to apply the analysis in this work to study the galaxy population-halo connection in other datasets such as the galaxy magnitude data and those from semi-analytical models, before applying the results to real observations. It is also worth extending these explorations to alternative halo mass definitions, given recently new understandings on the physical boundaries of halos such as the splashback radius~\citep[]{Diemer14,Adhikari14,Shi2016} and the depletion radius~\citep{FH21,LH21}.%It is also suggested that we may propose a conditional stellar mass function with member galaxies in the halo as parameters to construct a more accurate halo mass proxy, and thus place better constraints on the relation between the dark matter halos and galaxies.

%In future work, we plan to test whether the same conclusions can be reached in semi-analytic model data, and then apply them to observations such as SDSS. Following this, we will try to summarize a function to build a better galaxy-halo connection.
%\begin{acknowledgments}
\section*{acknowledgments}
We acknowledge helpful discussions with Wenting Wang, Rui Shi and Qingyang Li. JH benefited from discussions with Houjun Mo and many others at the assembly bias workshop at SJTU in 2019 which motivated this study. This work is supported by National Key Basic Research and Development Program of China (No.\ 2018YFA0404504), NSFC (11973032, 11890691, 11621303), 111 project (No.\ B20019), and the science research grants from the China Manned Space Project (No.\ CMS-CSST-2021-A03). The computation of this work is done on the \textsc{Gravity} supercomputer at the Department of Astronomy, Shanghai Jiao Tong University. 

The IllustrisTNG simulations were undertaken with compute time awarded by the Gauss Centre for Supercomputing (GCS) under GCS Large-Scale Projects GCS-ILLU and GCS-DWAR on the GCS share of the supercomputer Hazel Hen at the High Performance Computing Center Stuttgart (HLRS), as well as on the machines of the Max Planck Computing and Data Facility (MPCDF) in Garching, Germany.
%\end{acknowledgments}

\appendix

\section{MDI feature importance}\label{app:MDI}
The RF package we used can directly output a Mean Decrease Impurity (MDI) based feature importance, which can be used for feature selection in absence of feature interactions. The impurity funciton in regression is simply defined to be the MSE as in Equation~\eqref{eq:MSE}.
%Each split will reduce the weighted average of impurity in decision tree. 
In node $t$ which splits into two child nodes $t_1$ and $t_2$, the impurity decrease can be found as
\begin{align*}
\Delta I(j,s,t)=I(t)-p_1 I(t_1)-p_2 I(t_2),
\end{align*}
where $p_i$ = $\frac{N_{t_i}}{N_t}$ is the proportion of the sample in $t$ distributed into node $t_i$. Adding up the contributions to the total reduction from all the nodes that are divided with feature $j$, we get the importance for $j$ in a single decision tree
\begin{align*}
\mathrm{Imp}(j)=\sum_{t\in T_j} p(t)\Delta I(j, s, t),
\end{align*}
%Therefore, on an individual tree $m$, the importance of a certain feature $X_j$ can be expressed by adding up the weighted impurity decreases $p(t)\Delta I(s_t, t)$ for all nodes $t$ where $X_j$ is used:
where $p(t)=\frac{N_t}{N}$ is the fraction of the full sample found in $t$ and $T_j$ is the set of all the nodes divided with feature $j$. Finally, the importance of a feature $j$ in a RF is obtained by averaging over its importances in all the consistuting trees. With this definition, the summation of the feature importances from all the features is normalized to unity.

%\bibliography{sample631}{}
%\bibliographystyle{aasjournal}

%\bibliography{sample631}{}
\bibliographystyle{aasjournal}

\end{document}